\documentclass[a4paper,12pt]{article}

\usepackage{indentfirst}

\usepackage{amsfonts,amsmath,amssymb,bm,a4wide,graphicx,cite,makeidx,multicol}

\begin{document}

\title{Neutrinos and electrons in background matter: a new approach}

\author{Alexander Studenikin \footnote{studenik@srd.sinp.msu.ru },
   \\
   \small {\it Department of Theoretical Physics,}
   \\
   \small {\it Moscow State University,}
   \\
   \small {\it 119992 Moscow,  Russia }}
\date{}
\maketitle

\sloppy

\begin{abstract}
We present a rather powerful method in investigations of different phenomena
that can appear when neutrinos and electrons are moving in the background
matter. This method is based on the use of the modified Dirac equations for the
particles wave functions, in which the correspondent effective potentials that
account for the matter influence on particles are included. The developed
approach establishes a basis for investigation of different phenomena which can
arise when neutrinos and electrons move in dense media, including those
peculiar for astrophysical and cosmological environments.  The approach
developed is similar to the Furry representation of quantum electrodynamics,
widely used for description of particles interactions in the presence of
external electromagnetic fields, and it works when a macroscopic amount of the
background particles are confined within the scale of a neutrino or electron de
Broglie wave lengths. We consider the modified Dirac equations for neutrinos
(of both Dirac and Majorana types) and electrons having the standard model
interactions with the background matter, however generalization to extensions
of the standard model is just straightforward. To illustrate how the developed
method works, we elaborate the quantum theories of the spin light of neutrino
($SL\nu$) and spin light of electron ($SLe$) in matter.
\end{abstract}

\section{Introduction}

The neutrino is a very fascinating particle which has remained
under the focus of intensive investigations, both theoretical an
experimental, for a couple of decades. These studies have given
evidence of an ultimate relation between the knowledge on neutrino
properties and understanding of fundamentals of particle physics.
The birth of the neutrino (this particle was predicted by W.Pauli
in 1930) was due to an attempt to explain the continuous spectrum
of beta-particles and correspondently to find ``a way out for
saving the law of conservation of energy" \cite{Pauli57}. This new
particle, called at first the ``neutron" and then renamed the
``neutrino", was an essential counterpart of the first model for
the weak interactions (E.Fermi, 1934). The further important
milestones of particle physics, such as the parity nonconcervation
(T.Lee, C.Yang, 1956) and the $V-A$ model of the local weak
interactions (E.Sudarshan, R.Marshak, 1956; R.Feynman,
M.Gell-Mann, 1958), as well as the structure of the
Glashow-Weinberg-Salam {\it standard model}, were based on the
clarification of the specific properties of the neutrino. To
summarize the present status of the neutrino physics we quote
\cite{MohSmi_hp0603118}: ``Recent discovery of flavour conversion
of solar, atmospheric, reactor and accelerator neutrinos have
conclusively established that neutrinos have nonzero mass and they
mix among themselves much like quarks, providing the first
evidence of new physics beyond the {\it standard model}". Thus, it
has happened not once that a novel discovery in neutrino physics
stimulates a far-reaching consequences in the theory of particle
interactions.

The neutrino plays a crucial role in particle physics because it
is a ``tiny" particle. Indeed, the scale of neutrino mass is much
lower than ones of the charged fermions ($m_{\nu_{f}}<<m_f, \ \
f=e,\mu, \tau$). In addition to electric neutrality and very small
(if non-zero) magnetic moment, the weak interactions of neutrino
with other particles are really weak. That is a reason for the
neutrino to appear under the focus of researchers during the
latest stages of a particular particle physics paradigm evolution
when all of the ``principal" phenomena have been already observed
and theoretically described. That is why the neutrino has often
provided a kind of a bridge to ``new physics".

In general, the neutrino most easily demonstrates its properties
when it is subjected to the influence of extreme external
conditions which can be provided by the presence of dense matter
or strong external electromagnetic fields. A crucial importance of
external influence on neutrinos is vividly demonstrated by the
fact that the resonant neutrino flavour oscillations in matter
\cite{WolPRD78MikSmiYF85} has been proven to be the mechanism for
solving the solar neutrino problem. The resonant neutrino spin (or
spin-flavour) oscillations in matter under the influence of
external transversal magnetic field  \cite{Akh88LimMar88} have
also important consequences in astrophysics and cosmology (see,
for instance, \cite{Raf_book}). In the presence of matter a
neutrino dispersion relation is modified
\cite{ManPRD88,NotRaf88,NiePRD89}, in particular, it has a minimum
at nonzero momentum \cite{ChaZiaPRD88,PanPLB91-PRD92WeiKiePRD97}.
As it was shown in \cite{ManPRD88,NotRaf88,NiePRD89}, the standard
result for the MSW effect can be derived using a modified Dirac
equation for the neutrino wave function with a matter potential
proportional to the density being added. The problem of a neutrino
mass generation in different media was also studied
\cite{OraSemSmoPLB89,HaxZhaPRD91} on the basis of modified Dirac
equations for neutrinos. On the same basis spontaneous
neutrino-pair creation in matter was also studied
\cite{LoePRL90,KacPLB98,KusPosPLB02} (neutrino-pair creation in a
perturbative way has been recently discussed in
\cite{Koe04092549}).

The aim of this paper is to present a rather powerful method in
investigations of different phenomena that can appear when
neutrinos are moving in the background matter
\cite{StuTerPLB05,GriStuTerPLB05}. In addition, we also
demonstrate how this method can be applied to electrons moving in
background matter \cite{StuJPA06,GriShiStuTerTro_12LomCon}. The
approach developed establishes a basis for investigation of
different phenomena which can arise when neutrinos and electrons
move in dense media, including those peculiar for astrophysical
and cosmological environments.

 The method discussed is based on the use of the modified Dirac equations for the particles wave
functions, in which the correspondent effective potentials that
account for the matter influence on the particles are included. It
is similar to the Furry representation \cite{FurPR51} in quantum
electrodynamics, widely used for description of particles
interactions in the presence of external electromagnetic fields.

In quantum electrodynamics, as well as in other more general
models of interaction, there are two different approaches to the
problem of particles interactions in the presence of external
electromagnetic fields. Within the first approach, in the presence
of external field one has to substitute the four-potential of the
external field $A_{\mu} (x)$ by $A' _{\mu}(x)$ in accordance with
\begin{equation}
A'_{\mu} (x)= A_{\mu}+A_{\mu}^{ext} (x),
\end{equation}
where $A_{\mu}(x)$ is the quantized radiation field, and $A_{\mu}^{ext} (x)$ is
the potential of the external field that is a given function of the
four-coordinate $x$. The appropriate Feynman diagram techniques, after the
external field potential $ A_{\mu}^{ext} (x)$ being substituted into the
interaction Hamiltonian, can be developed. This approach implies that the
external filed is accounted for within the perturbation-theory expansion and
additional lines on the Feynman diagrams, which stand for interaction with the
external field, appears  \cite {BogShi80}. Note that applications of the
approach discussed is limited by the strength of the external field. For
instance, it can not be used in the case very strong electromagnetic fields or
in the case when non-liner over the external field effects are important.

There is another approach to the problem of particles interaction
in the presence of external electromagnetic fields which enables
one to account for the external field effect exactly, rather than
within the perturbation-series expansion discussed above. In this
techniques, known in quantum electrodynamics as the Furry
representation, the evolution operator $U_{F}(t_1, t_2)$, which
determines the matrix element of the process, is represented in
the usual form
\begin{equation}
U_{F} (t_1, t_2)= T exp \bigg[-i \int
\limits_{t_1}^{t_2}j_{\mu}(x) A^{\mu}{d}x  \bigg],
\end{equation}
where $A_{\mu}(x)$ is the quantized part of the potential corresponding to the
radiation field, which is accounted within the perturbation-series techniques.
At the same time, the charged particles current is represented in the form
\begin{equation}
j_{\mu}(x)={e \over 2}\big[\overline \Psi_F \gamma _{\mu}, \Psi_F
\big],
\end{equation}
where $\Psi_F$ are the exact solutions of the quantum wave
equation for the charged particles (that is the Dirac equation in
the case of an electron) in the presence of external
electromagnetic field given by the classical non-quantized
potential $A_{\mu}^{ext} (x)$. Note that within this approach the
interaction of charged particles with the external electromagnetic
field is taken into account exactly while the radiation field is
allowed for by perturbation-series expansion techniques. A
detailed discussion of this method can be found in
\cite{SokTerSynRad68}.

Note that our focus in this paper is on the standard model interactions of
neutrinos and electrons in the background matter (i.e., we consider the
standard model processes with participation of neutrinos and electrons that
proceed under the influence of the standard-model-interacting matter
background). It worth to be mentioned, that on the same ideological basis
(which implies the use of the exact solutions of the corresponding Dirac
equations) different processes with neutrinos and electrons beyond the standard
model (and also with non-standard model interactions with an environment) were
considered in the literature
\cite{BerVysBerSmiPLB87_89GiuKimLeeLamPRD92GiuKimLeeLamPRD92BerRosPLB94,
CollKosPRD98,ZhuLobMurPRD06}.

Here below we start with the discussion on the form of the modified Dirac
equation for the neutrino wave function in the presence of the background
matter. We consider the two possibilities that are provided by the use of the
modified Dirac-Pauli and modified Dirac equations. For both cases we explicitly
evaluate the exact solutions for the wave functions and derive the neutrino
energy spectra in the background matter. Then we also consider the particular
case of the Majorana neutrino and determine the energy spectrum for this case.
For the Dirac neutrino, using the obtained neutrino energy spectrum, we discuss
several interesting phenomena, such as neutrino trapping, neutrino reflection
and neutrino-antineutrino pair production and annihilation that can appear at
the boundary of media with different densities. Then we demonstrate how the
evaluated neutrino quantum states in matter (the exact neutrino wave functions
and energy spectrum in matter) can be used for investigation of different
processes in the presence of matter. As an example, we develop the quantum
theory of the spin light of neutrino ($SL\nu$), a new type of the
electromagnetic radiation that can be emitted by the neutrino moving in the
background matter. In the final section of the paper, we derive the modified
Dirac equation for an electron moving in the background matter and obtain in
explicit form the electron wave function and energy spectrum. On this bases we
predict a new type of radiation that can be emitted by the electron moving in
matter (we have termed this radiation as ``spin light of electron" in matter
\cite{StuJPA06,GriShiStuTerTro_12LomCon}).

\section{Neutrino wave function and energy spectrum in matter}

In this section we discuss two quantum equations, the modified Dirac-Pauli and
modified Dirac ones, for a massive neutrino wave function in the background
matter. The two wave functions and energy spectra, that correspond to these two
equations, in general do not coincide. However, for the relativistic neutrino
and in the linear approximation over the matter density the two energy spectra
lead to equal results for the energy difference of the left-handed and
right-handed neutrino states. So that, in the lowest approximation over the
matter density, the results for the $SL\nu$ photon energy, rate and power
obtained on the basis of the two equations discussed  (see Section 3) are
equal.

\subsection{Modified Dirac-Pauli equation for neutrino in
matter}

To derive the quantum equation for a neutrino wave function in the background
matter we start with the well-known Dirac-Pauli equation for a neutral fermion
with non-zero magnetic moment. For a massive neutrino moving in an
electromagnetic field $F_{\mu \nu}$ this equation is given by
\begin{equation}\label{D_P}
\Big( i\gamma^{\mu}\partial_{\mu} - m -\frac{\mu}{2}\sigma ^{\mu
\nu}F_{\mu \nu}\Big)\Psi(x)=0,
\end{equation}
where $m$ and $\mu$ are the neutrino mass and magnetic moment
\cite{FujShr80} \footnote{For the recent studies of a massive
neutrino electromagnetic properties, including discussion on the
neutrino magnetic moment, see Ref.\cite{DvoStuPRD_JETP04}},
$\sigma^{\mu \nu}=i/2 \big(\gamma^{\mu }\gamma^{\nu}-\gamma^{\nu}
\gamma^{\mu}\big)$. It worth to be noted here that Eq.({\ref{D_P})
can be obtained in the linear approximation over the
electromagnetic field from the Dirac-Schwinger equation, which in
the case of the neutrino takes the following form
\cite{BorZhuTer88}:
\begin{equation}\label{D_S}
(i\gamma^\mu\partial_\mu-m) \Psi (x) =\ \int M^{F}(x',x)\Psi
(x')dx' ,
\end{equation}
where  $M^{F}(x',x)$ is the neutrino mass operator in the presence
of the external electromagnetic field.

For the case of the external magnetic filed, the Hamiltonian form
of the equation (\ref{D_P}) reads
\begin{equation}\label{H_form}
i\frac{\partial}{\partial t}\Psi({\bf r},t)=\hat H_{F}\Psi({\bf
r},t),
\end{equation}
where
\begin{equation}\label{H}
  \hat H_{F}=\hat {\bm{\alpha}} {\bf p} + \hat {\beta}m + \hat V_F,
  \hat V_F= -\mu \hat {\beta} {\bf {\hat \Sigma}} {\bf B},
\end{equation}
and ${\bf B}$ is the magnetic field vector. We use the Pauli-Dirac
representation of the Dirac matrices $\hat {\bm \alpha}$ and $\hat
{\beta}$, in which
\begin{equation}\label{a_b}
    \hat {\bm \alpha}=
\begin{pmatrix}{0}&{\hat {\bm \sigma}} \\
{\hat {\bm \sigma}}& {0}
\end{pmatrix}=\gamma_0{\bm \gamma}, \ \ \
\hat {\beta}=\begin{pmatrix}{1}&{0} \\
{0}& {-1}
\end{pmatrix}=\gamma_0, \ \ {\hat {\bm \Sigma}}=
\begin{pmatrix}{\hat {\bm \sigma}}&{0} \\
{0}&{\hat {\bm \sigma}}
\end{pmatrix},
\end{equation}
where ${\hat {\bm \sigma}}=({ \sigma}_{1},{ \sigma}_{2},{
\sigma}_{3})$ and  $\sigma$ denotes the Pauli matrixes.

 Recently in a series of our papers
\cite{EgoLobStuPLB00LobStuPLB01,DvoStuJHEP02,LobStuPLB03_04DvoGriStuIJMPD05,StuPAN04}
we have developed the quasi-classical approach to the massive neutrino spin
evolution in the presence of external fields and background matter. In
particular, we have shown that the well known Bargmann-Michel-Telegdi (BMT)
equation \cite{BarMicTelPRL59} of the electrodynamics can be generalized for
the case of a neutrino moving in the background matter and being also under the
influence of external electromagnetic fields. The proposed new equation for a
neutrino, which simultaneously accounts  for the electromagnetic interaction
with external fields and also for the weak interaction with particles of the
background matter, was obtained from the BMT equation by the following
substitution of the electromagnetic field tensor $F_{\mu\nu}=(\bf E,\bf B)$:
\begin{equation}\label{sub}
F_{\mu\nu} \rightarrow E_{\mu\nu}= F_{\mu\nu}+G_{\mu\nu},
\end{equation}
where the tensor $G_{\mu\nu}=(-{\bf P},{\bf M})$ accounts for the
neutrino interactions with particles of the environment. The
substitution (\ref{sub}) implies that in the presence of matter
the magnetic $\bf B$ and electric $\bf E$ fields are shifted by
the vectors $\bf M$ and $\bf P$, respectively:
\begin{equation}
\bf B \rightarrow \bf B +\bf M, \ \ \bf E \rightarrow \bf E - \bf
P. \label{11}
\end{equation}
We have also shown  how to construct the tensor $G_{\mu \nu}$ with
the use of the neutrino speed, matter speed, and matter
polarization four-vectors.

Now let us consider the case of a neutrino moving in matter
without any electromagnetic field in the background. The quantum
equation for the neutrino wave function can be obtained from
(\ref{D_P}) with application of the substitution (\ref{sub}) which
now becomes

\begin{equation}\label{sub_1}
  F_{\mu \nu}\rightarrow G_{\mu \nu}.
\end{equation}
Thus, we get the quantum equation for the neutrino wave function
in the presence of the background matter in the form
\cite{GriStuTer_11LomCon}
\begin{equation}\label{D_P_matter}
\Big( i\gamma^{\mu}\partial_{\mu} - m -\frac{\mu}{2}\sigma ^{\mu
\nu}G_{\mu \nu}\Big)\Psi(x)=0,
\end{equation}
that can be regarded as the modified Dirac-Pauli equation. The
generalization of this equation for the case when an
electromagnetic field is present in the environment, in addition
to the background matter, is just straightforward. In particular,
the case of external magnetic field is considered below.

The detailed discussion on the evaluation of the tensor $G_{\mu
\nu}$ is given in
\cite{EgoLobStuPLB00LobStuPLB01,DvoStuJHEP02,StuPAN04}. We
consider here, for simplicity,  the case of the unpolarized matter
composed of the only one type of fermions  of a constant density.
For a background of only electrons we  get
\begin{equation}\label{G_}
G^{\mu \nu}= \gamma \rho^{(1)} n
\begin{pmatrix}{0}&{0}& {0}&{0} \\
{0}& {0}& {-\beta_{3}}&{\beta_{2}} \\
{0}&{\beta_{3}}& {0}&{-\beta_{1}} \\
{0}&{-\beta_{2}}& {\beta_{1}}& {0}
\end{pmatrix}, \gamma=(1-{\bm\beta}^{2})^{-1/2},
\rho^{(1)}=\frac{\tilde{G}_{F}}{2\sqrt{2}\mu},
\tilde{G}_{F}=G_F(1+4\sin^{2}\theta_{W}),
\end{equation}
where ${\bm \beta}=(\beta_{1},\beta_{2},\beta_{3})$ is the neutrino
three-dimensional speed, $n$ denotes the number density of the background
electrons. From (\ref{G_}) and the two equations, (\ref{D_P}) and
(\ref{D_P_matter}), it is possible to see that the term $\gamma \rho^{(1)}n{\bm
\beta}$ in (\ref{D_P_matter}) plays the role of the magnetic field $\bf B$ in
(\ref{D_P}). Therefore, the Hamiltonian form of (\ref{D_P_matter}) is
\begin{equation}\label{H_G_form}
i\frac{\partial}{\partial t}\Psi({\bf r},t)=\hat H_{G}\Psi({\bf
r},t),
\end{equation}
where
\begin{equation}\label{H_G_1}
  \hat H_{G}=\hat {\bm{\alpha}} {\bf p} + \hat {\beta}m + \hat V_G,
\end{equation}
and
\begin{equation}\label{V_G}
\hat V_G= -\mu  \frac{\rho^{(1)} n}{m}
  \hat {\beta}{\bf \Sigma}{\bf p},
\end{equation}
here $\bf p$ is the neutrino momentum. From (\ref{V_G}) it is just
straightforward that the potential energy in matter depends on the
neutrino helicity.

The form of the Hamiltonian (\ref{H_G_1}) ensures that the
operators of the momentum, $\hat {\bf p}$, and helicity, ${\bf
\Sigma} {\bf p}/p$, are integrals of motion. That is why for the
stationary states we can write
\begin{equation}\label{stat_states}
\Psi({\bf r},t)=e^{-i( Et-{\bf p}{\bf r})}u({\bf p},E),
\ \ \ u({\bf p},E)=\begin{pmatrix}{\varphi}\\
{\chi}
\end{pmatrix},
\end{equation}
where $u({\bf p},E)$ is independent on the coordinates and time
and can be expressed in terms of the two-component spinors
$\varphi$ and $\chi$. Substituting (\ref{stat_states}) into
Eq.(\ref{H_G_form}), we get the two equations
\begin{equation}\label{H_u}
    ({\bm \sigma}{\bf p})\chi-\big(E-m+\alpha({\bm \sigma}{\bf
    p})\big)\varphi=0, \ \ \
({\bm \sigma}{\bf p})\varphi-\big(E-m+\alpha({\bm \sigma}{\bf
    p})\big)\chi=0.
\end{equation}
Suppose that $\varphi$ and $\chi$ satisfy the following equations,
\begin{equation}\label{helicity}
  ({\bm \sigma}{\bf p})\varphi=sp\varphi, \ \
  ({\bm \sigma}{\bf p})\chi=sp\chi,
\end{equation}
where $s=\pm 1$ specify the two neutrino helicity states. Upon the
condition that the set of Eqs.(\ref{H_u}) has a non-trivial
solution, we arrive to the energy spectrum of a neutrino moving in
the background matter:
\begin{equation}\label{Energy_1}
  E={\sqrt{{\bf p}^{2}(1+\alpha^{2})+m^2 -2\alpha m p s}},
\ \  \ \   \alpha=\frac{\mu \rho^{(1)}}{m}n=
  \frac{1}{2\sqrt{2}}{\tilde G}_{F}\frac{n}{m}.
\end{equation}
 It is important that the the neutrino energy in the
background matter depends on the state of the neutrino
longitudinal polarization (helicity), i.e. the left-handed and
right-handed neutrinos with equal momentum have different
energies.

The obtained expression  (\ref{Energy_1}) for the neutrino energy
can be transformed to the form
\begin{equation}\label{Energy_2}
E=\sqrt{{\bf p}^{2}+m^2\Big(1-s\frac{\alpha p}{m}\Big)^{2}}.
\end{equation}
It is easy to see that the energy spectrum of a neutrino in
vacuum, which is derived on the basis of the Dirac equation, is
modified in the presence of matter by the formal shift of the
neutrino mass
\begin{equation}\label{substitution}
  m\rightarrow m\Big(1-s\frac{\alpha p}{m}\Big).
\end{equation}

The procedure, similar to one used for the derivation of the
solution of the Dirac equation in vacuum, can be adopted for the
case of the neutrino moving in matter. We apply this procedure to
equation (\ref{H_G_form}) and arrive to the final form of the wave
function of a neutrino moving in the background matter:
\begin{equation}\label{wave_function_1}
\Psi_{{\bf p},s}({\bf r},t)=\frac{e^{-i( Et-{\bf p}{\bf
r})}}{2L^{\frac{3}{2}}}
\begin{pmatrix}{\sqrt{1+ \frac{m-s\alpha p}{E}}}
\ \sqrt{1+s\frac{p_{3}}{p}}
\\
{s \sqrt{1+ \frac{m-s\alpha p}{E}}} \ \sqrt{1-s\frac{p_{3}}{p}}\ \
e^{i\delta}
\\
{  s\sqrt{1- \frac{m-s\alpha p}{E}}} \ \sqrt{1+s\frac{p_{3}}{p}}
\\
{\sqrt{1- \frac{m-s\alpha p}{E}}} \ \ \sqrt{1-s\frac{p_{3}}{p}}\
e^{i\delta}
\end{pmatrix} ,
\end{equation}
where $L$ is the normalization length and $ \delta=\arctan p_{y}/p_{x}$. In the
limit of vanishing matter density, when $\alpha\rightarrow 0$, the wave
function of (\ref{wave_function_1}) transforms to the vacuum solution of the
Dirac equation.

Calculations on the basis of the modified Dirac-Pauli equation
(\ref{D_P_matter}) enables us to reproduce, to the lowest order of
the expansion over the matter density, the correct energy
difference between the two neutrino helicity states in matter
\footnote{ Note that in the relativistic energy limit the
negative-helicity neutrino state is dominated by the left-handed
chiral state ($\nu_{-}\approx \nu_{L}$), whereas the
positive-helicity state is dominated by the right-handed chiral
state ($\nu_{+}\approx \nu_{R}$)}. On the basis of the obtained
neutrino energy spectrum (\ref{Energy_1}), in the low matter
density limit $\alpha\frac {pm}{E_{0}^{2}} \ll 1$, we can
reproduce the correct result for the energy difference $\Delta
E=E(s=-1)-E(s=+1)$ of the two neutrino helicity states:
\begin{equation}\label{delta_Energy}
  \Delta E\approx 2m\alpha \frac{p}{E_{0}},
\end{equation}
where we use the notation $E_0=\sqrt{p^2 +m^2}$. Therefore, for the
relativistic neutrinos  one can derive, using (\ref{V_G}), the probability of
the neutrino spin oscillations $\nu_{L} \leftrightarrow \nu_{R}$ in transversal
magnetic field with the correct form of the matter term \cite{Akh88LimMar88}
(for the further details see Section 2.2).

\subsection{Modified Dirac-Pauli equation in magnetized and
polarized matter}

We should like to note that it is possible to generalize the Dirac-Pauli
equation (\ref{D_P}) (or (\ref{D_P_matter})) for the case when a neutrino is
moving in the magnetized background matter. For this case (i.e., when the
effects of matter and magnetic field on neutrino have to be accounted for) the
modified Dirac-Pauli equation is \cite{GriStuTer_11LomCon}
\begin{equation}\label{D_P_matter_1}
\Big\{ i\gamma^{\mu}\partial_{\mu} - m -\frac{\mu}{2}\sigma ^{\mu
\nu}(F_{\mu \nu}+G_{\mu \nu})\Big\}\Psi(x)=0,
\end{equation}
where the magnetic field $\bf B$ enters through the tensor
$F_{\mu\nu}$. The neutrino energy in the magnetized matter can be
obtained from (\ref{Energy_1}) by the following redefinition
\begin{equation}\label{redefin}
  \alpha \rightarrow \alpha '=\alpha
  +\frac{\mu B_{\parallel}}{p},
\end{equation}
where $B_{\parallel}= ({\bf B} {\bf p})/ p$. Thus, the neutrino
energy in this case reads
\begin{equation}\label{energy_2}
E=\sqrt{{\bf p}^{2}+m^2\Big(1-s\frac{\alpha p +\mu
B_{\parallel}}{m}\Big)^{2}}.
\end{equation}
For the relativistic neutrinos the expression of
Eq.(\ref{energy_2}) gives, in the linear approximation over the
matter density and the magnetic field strength, the correct value
(see \cite{EgoLobStuPLB00LobStuPLB01,StuPAN04}) for the energy
difference of the two opposite helicity states in the magnetized
matter:
\begin{equation}\label{Delta}
\Delta_{eff}= {{\tilde {G}}_F \over \sqrt{2}}n +2{\mu
B_{\parallel} \over \gamma}.
\end{equation}
On this basis we can also consider the neutrino spin oscillations
in the presence of non-moving matter being under the influence of
an arbitrary constant magnetic field ${\bf B}={\bf B}_{\parallel}
+{\bf B}_{\perp}$, here $\bf B_{\perp}$ is the transversal to the
neutrino momentum component of the external field. In the
adiabatic approximation the probability of the oscillations $\nu_L
\leftrightarrow \nu_R$ can be written in the form,
\begin{equation}\label{P_L_R}
P_{\nu_L \rightarrow \nu_R} (x)=\sin^{2} 2\theta_{eff}
\sin^{2}{\pi x \over L_{eff}},\end{equation}
\begin{equation}\label{sin}
sin^{2} 2\theta_{eff}={E^2_{eff} \over
{E^{2}_{eff}+\Delta^{2}_{eff}}}, L_{eff}={2\pi \over
\sqrt{E^{2}_{eff}+\Delta^{2}_{eff}}},
\end{equation}
where $E_{eff}=2\mu B_{\perp}$ (terms $\sim \gamma^{-1}$ are
omitted  here), and $x$ is the distance travelled by the neutrino.
The obtained expression for the oscillation probability confirms
our previous result of refs.\cite{EgoLobStuPLB00LobStuPLB01,
StuPAN04}.

Let us now shortly discuss the effect of matter polarization. Consider the case
of matter composed of electrons in the presence of such strong background
magnetic field so that the following condition is valid
\begin{equation}\label{str_B}
  B>\frac{p_{F}^{2}}{2e},
\end{equation}
where $p_{F}=\sqrt{\mu ^2 - m^{2}_{e}}$,  $\mu$ and $m_e$ are, respectively,
the Fermi momentum, chemical potential, and mass of electrons. Then all of the
electrons occupy the lowest Landau level (see, for instance,
\cite{NunSemSmiValNPB97}), therefore the matter is completely polarized in the
direction opposite to the unit vector $\frac{{\bf B}}{B}$. From the general
expression for the tensor $G_{\mu \nu}$ (see the second paper of
\cite{EgoLobStuPLB00LobStuPLB01}) we get
\begin{equation}\label{G_1}
G^{\mu \nu}= \gamma  n \left\{
\rho^{(1)}\begin{pmatrix}{0}&{0}& {0}&{0} \\
{0}& {0}& {-\beta_{3}}&{\beta_{2}} \\
{0}&{\beta_{3}}& {0}&{-\beta_{1}} \\
{0}&{-\beta_{2}}& {\beta_{1}}& {0}
\end{pmatrix}
+
\rho^{(2)}\begin{pmatrix}{0}&{-\beta_{2}}& {\beta_{1}}&{0} \\
{\beta_{2}}& {0}& {-\beta_{0}}&{0} \\
{-\beta_{1}}&{\beta_{0}}& {0}&{0} \\
{0}&{0}& {0}& {0}
\end{pmatrix}\right\},
\end{equation}
\begin{equation*}
\rho^{(2)}=-\frac{{G}_{F}}{2\sqrt{2}\mu}.
\end{equation*}
Thus, the modified Dirac-Pauli equation (\ref{D_P_matter_1}) with the tensor
$G_{\mu \nu}$ given by (\ref{G_1}) can be used for description of the neutrino
motion in magnetized and totally polarized (parallel or anti-parallel to the
magnetic field vector $\bm B$ direction) matter. The neutrino energy in such a
case can be obtained from (\ref{Energy_1}) by the following redefinition
\begin{equation}\label{redefin}
  \alpha \rightarrow {\tilde \alpha} =\alpha
  \left[1- \frac{sign \left(\frac{B_{\parallel}}{B}\right)}
  {1+\sin^{2}
  4\theta_{W}}\right]
  +\frac{\mu B_{\parallel}}{p}.
\end{equation}
In Eq.(\ref{redefin}), the second term in brackets  accounts for the effect of
the matter polarization. It follows, that the effect of the matter polarization
can reasonably change the total matter contribution to the neutrino energy
(\ref{energy_2}) (see also \cite{NunSemSmiValNPB97}).

\section{Modified Dirac equation for neutrino in matter}

 In \cite{StuTerPLB05}  (see  also
\cite{GriStuTerPLB05,StuJPA06}) we derived the modified Dirac
equation for the neutrino wave function exactly accounting for the
neutrino interaction with matter. Let us consider the case of
matter composed of electrons, neutrons, and protons and also
suppose that the neutrino interaction with background particles is
given by the standard model supplied with the singlet right-handed
neutrino. The corresponding addition to the effective interaction
Lagrangian is given by
\begin{equation}\label{Lag_f}
\Delta L_{eff}=-f^\mu \Big(\bar \nu \gamma_\mu {1+\gamma^5 \over
2} \nu \Big), \ \ f^\mu={\sqrt2 G_F }\sum\limits_{f=e,p,n}
j^{\mu}_{f}q^{(1)}_{f}+\lambda^{\mu}_{f}q^{(2)}_{f},
\end{equation}
where
\begin{equation}\label{q_f}
q^{(1)}_{f}=
(I_{3L}^{(f)}-2Q^{(f)}\sin^{2}\theta_{W}+\delta_{ef}), \
q^{(2)}_{f}=-(I_{3L}^{(f)}+\delta_{ef}), \  \delta_{ef}=\left\{
\begin{tabular}{l l}
1 & for {\it f=e}, \\
0 & for {\it f=n, p}. \\
\end{tabular}
\right.
\end{equation}
Here $I_{3L}^{(f)}$ and $Q^{(f)}$ are, respectively, the values of
the isospin third components and electric charges of the particles
of matter ($f=e,n,p$). The corresponding currents $j_{f}^{\mu}$
and polarization vectors $\lambda_{f}^{\mu}$ are
\begin{equation}\label{j}
j_{f}^\mu=(n_f,n_f{\bf v}_f),
\ \ \ \lambda_f^{\mu} =\Bigg(n_f ({\bm \zeta}_f {\bf v}_f ), n_f
{\bm \zeta}_f \sqrt{1-v_f^2}+ {{n_f {\bf v}_f ({\bm \zeta}_f {\bf
v}_f )} \over {1+\sqrt{1- v_f^2}}}\Bigg),
\end{equation}
where $\theta _{W}$ is the Weinberg angle. In the above formulas (\ref{j}),
$n_f$, ${\bf v}_f$ and ${\bm \zeta}_f \ (0\leq |{\bm \zeta}_f |^2 \leq 1)$
stand, respectively, for the invariant number densities, average speeds and
polarization vectors of the matter components. A detailed discussion on the
meaning of these characteristics can be found in
\cite{EgoLobStuPLB00LobStuPLB01,DvoStuJHEP02,
LobStuPLB03_04DvoGriStuIJMPD05,StuPAN04}. Using the standard model Lagrangian
with the extra term (\ref{Lag_f}), we derive the modified Dirac equation for
the neutrino wave function in matter:
\begin{equation}\label{new} \Big\{
i\gamma_{\mu}\partial^{\mu}-\frac{1}{2}
\gamma_{\mu}(1+\gamma_{5})f^{\mu}-m \Big\}\Psi(x)=0.
\end{equation}
This is the most general form of the equation for the neutrino
wave function in which the effective potential
$V_{\mu}=\frac{1}{2}(1+\gamma_{5})f_{\mu}$ includes both the
neutral and charged current interactions of the neutrino with the
background particles and which could also account for effects  of
matter motion and polarization. It should be mentioned that other
modifications of the Dirac equation were previously used in
\cite{ManPRD88}-\cite{HaxZhaPRD91} for studies of the neutrino
dispersion relations, neutrino mass generation and neutrino
oscillations in the presence of matter. Note that the
corresponding quantum wave equation for a Majorana neutrino can be
obtained from (\ref{new}) via the substitution $1+\gamma _5
\rightarrow 2\gamma _5$ (see Section 3.1 below and also
\cite{GriStuTerNANP_PAN06,PivStuPOS05}).

In the further discussion below we consider the case when no
electromagnetic field is present in the background. We also
suppose that the matter is unpolarized, $\lambda^{\mu}=0$.
Therefore, the term describing the neutrino interaction with the
matter is given by
\begin{equation}\label{f}
f^\mu=\frac{\tilde{G}_{F}}{\sqrt2}(n,n{\bf v}),
\end{equation}
where we use the notation $\tilde{G}_{F}={G}_{F}(1+4\sin^2 \theta
_W)$.

In the rest frame of the matter the Hamiltonian form of the equation
(\ref{new}) can be written as
\begin{equation}\label{H_matter}
i\frac{\partial}{\partial t}\Psi({\bf r},t)=\hat H_{matt}\Psi({\bf
r},t),
\end{equation}
where
\begin{equation}\label{H_G}
  \hat H_{matt}=\hat {\bm{\alpha}} {\bf p} + \hat {\beta}m +
  \hat V_{matt},
\end{equation}
and
\begin{equation}\label{V_matt}
\hat V_{matt}= \frac{1}{2\sqrt{2}}(1+\gamma_{5}){\tilde {G}}_{F}n.
\end{equation}
The form of the Hamiltonian (\ref{H_G}) implies that the operators of the
momentum, $\hat {\bf p}$, and longitudinal polarization, ${\hat{\bf \Sigma}}
{\bf p}/p$, are the integrals of motion. So that, in particular, we have
\begin{equation}\label{helicity}
  \frac{{\hat{\bf \Sigma}}{\bf p}}{p}
  \Psi({\bf r},t)=s\Psi({\bf r},t),
 \ \ {\hat {\bm \Sigma}}=
\begin{pmatrix}{\hat {\bm \sigma}}&{0} \\
{0}&{\hat {\bm \sigma}}
\end{pmatrix},
\end{equation}
where the values $s=\pm 1$ specify the two neutrino helicity
states, $\nu_{+}$ and  $\nu_{-}$. In the relativistic limit the
negative-helicity neutrino state is dominated by the left-handed
chiral state ($\nu_{-}\approx \nu_{L}$), whereas the
positive-helicity state is dominated by the right-handed chiral
state ($\nu_{+}\approx \nu_{R}$).

For the stationary states of the equation (\ref{new}) we get
\begin{equation}\label{stat_states}
\Psi({\bf r},t)=e^{-i(  E_{\varepsilon}t-{\bf p}{\bf r})}u({\bf
p},E_{\varepsilon}),
\end{equation}
where $u({\bf p},E_{\varepsilon})$ is independent on the
coordinates and time. Upon the condition that the equation
(\ref{new}) has a non-trivial solution, we arrive to the energy
spectrum of a neutrino moving in the background matter:
\begin{equation}\label{Energy}
  E_{\varepsilon}=\varepsilon{\sqrt{{\bf p}^{2}\Big(1-s\alpha \frac{m}{p}\Big)^{2}
  +m^2} +\alpha m} ,
\end{equation}
where we use the notation
\begin{equation}\label{alpha}
  \alpha=\frac{1}{2\sqrt{2}}{\tilde G}_{F}\frac{n}{m}.
\end{equation}
The quantity $\varepsilon=\pm 1$ splits the solutions into the two
branches that in the limit of the vanishing matter density,
$\alpha\rightarrow 0$, reproduce the positive and
negative-frequency solutions, respectively. It is also important
to note that the neutrino energy in the background matter depends
on the state of the neutrino longitudinal polarization, i.e. in
the relativistic case the left-handed and right-handed neutrinos
with equal momenta have different energies.

We get the exact solution of the modified Dirac equation in the
form \cite{StuTerPLB05}
\begin{equation}\label{wave_function}
\Psi_{\varepsilon, {\bf p},s}({\bf r},t)=\frac{e^{-i(
E_{\varepsilon}t-{\bf p}{\bf r})}}{2L^{\frac{3}{2}}}
\begin{pmatrix}{\sqrt{1+ \frac{m}{ E_{\varepsilon}-\alpha m}}}
\ \sqrt{1+s\frac{p_{3}}{p}}
\\
{s \sqrt{1+ \frac{m}{ E_{\varepsilon}-\alpha m}}} \
\sqrt{1-s\frac{p_{3}}{p}}\ \ e^{i\delta}
\\
{  s\varepsilon\sqrt{1- \frac{m}{ E_{\varepsilon}-\alpha m}}} \
\sqrt{1+s\frac{p_{3}}{p}}
\\
{\varepsilon\sqrt{1- \frac{m}{ E_{\varepsilon}-\alpha m}}} \ \
\sqrt{1-s\frac{p_{3}}{p}}\ e^{i\delta}
\end{pmatrix},
\end{equation}
 where the energy $E_{\varepsilon}$ is given by
(\ref{Energy}), $L$ is the normalization length and
$\delta=\arctan{p_2/p_1}$. In the limit of vanishing density of
matter, when $\alpha\rightarrow 0$, the wave function
(\ref{wave_function}) transforms to the vacuum solution of the
Dirac equation.

The quantum equation (\ref{new}) for a neutrino in the background
matter with the obtained exact solution (\ref{wave_function}) and
energy spectrum (\ref{Energy}) establish a basis for a very
effective method (similar to the Furry representation of quantum
electrodynamics) in investigations of different phenomena that can
appear when neutrinos are moving in the media.

Let us now consider in some detail the properties of a neutrino energy spectrum
(\ref{Energy}) in the background matter that are very important for
understanding of the mechanism of the neutrino spin light phenomena. For the
fixed magnitude of the neutrino momentum $p$ there are the two values for the
``positive sign" ($\varepsilon =+1$) energies
\begin{equation}\label{Energy_nu}
  E^{s=+1}={\sqrt{{\bf p}^{2}\Big(1-\alpha \frac{m}{p}\Big)^{2}
  +m^2} +\alpha m}, \ \ \
 E^{s=-1}={\sqrt{{\bf p}^{2}\Big(1+\alpha \frac{m}{p}\Big)^{2}
  +m^2} +\alpha m},
\end{equation}
that determine the positive- and negative-helicity eigenstates, respectively.
The energies in (\ref{Energy_nu}) correspond to the particle (neutrino)
solutions in the background matter. The two other values for the energy,
corresponding to the negative sign $\varepsilon =-1$, are for the antiparticle
solutions. As usual, by changing the sign of the energy, we obtain the values
\begin{equation}\label{Energy_anti_nu}
  {\tilde E}^{s=+1}={\sqrt{{\bf p}^{2}
  \Big(1-\alpha \frac{m}{p}\Big)^{2}
  +m^2} -\alpha m}, \ \ \
  {\tilde E}^{s=-1}={\sqrt{{\bf p}^{2}
  \Big(1+\alpha \frac{m}{p}\Big)^{2}
  +m^2} -\alpha m},
\end{equation}
that correspond to the positive- and negative-helicity antineutrino states in
the matter. The expressions in (\ref{Energy_nu}) and (\ref{Energy_anti_nu})
would reproduce the neutrino dispersion relations of
\cite{PanPLB91-PRD92WeiKiePRD97}, if the contribution of the neutral-current
interaction to the neutrino potential were left out.

Note that on the basis of the obtained energy spectrum
(\ref{Energy}) the neutrino trapping and reflection, the
neutrino-antineutrino pair annihilation and creation in a medium
can be studied
\cite{ChaZiaPRD88,LoePRL90,PanPLB91-PRD92WeiKiePRD97}. In the
general case of matter composed of electrons, neutrons and protons
the matter density parameter $\alpha$ for different neutrino
species is
\begin{equation}\label{alpha}
  \alpha_{\nu_e,\nu_\mu,\nu_\tau}=
  \frac{1}{2\sqrt{2}}\frac{G_F}{m}\Big(n_e(4\sin^2 \theta
_W+\varrho)+n_p(1-4\sin^2 \theta _W)-n_n\Big),
\end{equation}
where $\varrho=1$ for the electron neutrino and $\varrho=-1$ for
the muon and tau neutrinos.

The analysis of the obtained energy spectrum (\ref{Energy_nu}),
(\ref{Energy_anti_nu}) enables us to predict some interesting phenomena that
may appear at the interface of the two media with different densities and, in
particular, at the interface between matter and vacuum. Indeed, as it follows
from (\ref{Energy_nu}) and (\ref{Energy_anti_nu}) (see also
\cite{GriStuTerNANP_PAN06}), the band-gap for neutrino and antineutrino in
matter is displaced with respect to the vacuum case in neutrino mass and is
determined by the condition $\alpha m- m\leq E<\alpha m + m$. For instance, let
us consider the case when there is no band-gap overlapping (it is possible for
$\alpha
>2$). This situation is illustrated in Fig.1.
\begin{figure}[h]\label{Reflaction}
\begin{center}
\includegraphics[width=0.7\textwidth]{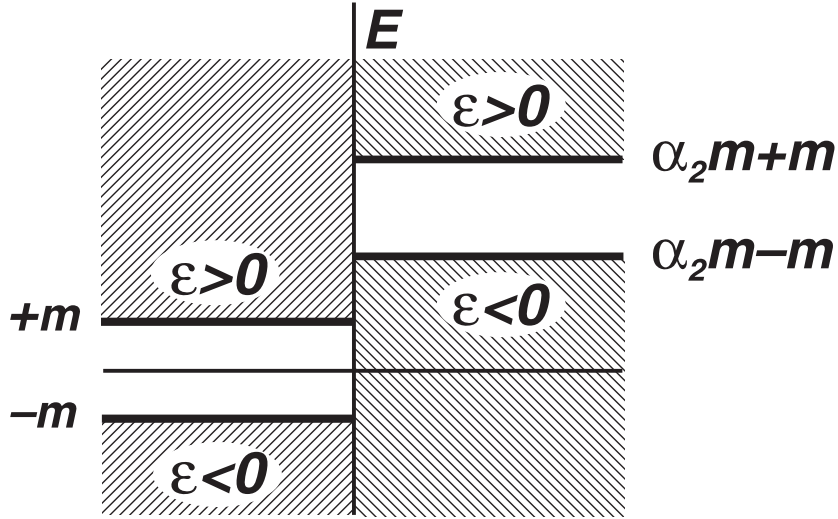}\\
\parbox{0.75\textwidth}{\caption{The interface between the vacuum (left-hand side of the picture)
    and the matter (right-hand side of the picture) with the corresponding
    neutrino band-gaps shown. The parameter $\alpha >2$. }}
\end{center}
\end{figure}
Let us consider first a neutrino moving in the vacuum towards the
interface with energy that falls into the band-gap region in
matter. In this case the neutrino has no chance to survive in the
matter and thus it is reflected from the interface. The same
situation is realized for the antineutrino moving in the matter
with energy falling into the band-gap in the vacuum. In this case
the antineutrino is trapped by the matter. When the energies of
neutrino in the vacuum or antineutrino in the medium fall into the
region between the two band-gaps the effects of the
neutrino-antineutrino annihilation or pair creation may occur
(see, for example, the first paper of Ref.
\cite{PanPLB91-PRD92WeiKiePRD97} and also
\cite{LoePRL90,KacPLB98,KusPosPLB02}).

\subsection{Majorana neutrino}

We have considered so far the case of Dirac neutrino. Now let us turn to
Majorana neutrino \cite{GriStuTerNANP_PAN06}. For a Majorana neutrino we derive
the following contribution to the effective Lagrangian accounting for the
interaction with the background medium
\begin{equation}\label{Lag_f_Majorana}
\Delta L_{eff}=-f^\mu (\bar \nu \gamma_\mu \gamma^5 \nu ),
\end{equation}
which leads to the Dirac equation
\begin{equation}\label{Dirac_Majorana} \Big\{
i\gamma_{\mu}\partial^{\mu}-\gamma_{\mu}\gamma_{5}f^{\mu}-m \Big\}\Psi(x)=0.
\end{equation}
This equation differs from the one, obtained in the Dirac case, by doubling of the interaction term
and lack of the vector part. The corresponding energy spectrum for the equation
(\ref{Dirac_Majorana}) is:
\begin{equation}\label{Energy_Majorana}
  E_{\varepsilon}=\varepsilon{\sqrt{{\bf p}^{2}\Big(1-2s\alpha \frac{m}{p}\Big)^{2}
  +m^2}}.
\end{equation}
From this expression it is clear, that the energy of the Majorana neutrino has
its minimal value equal to the neutrino mass, $E =m$. This means that no
effects are anticipated for the Majorana neutrino such as the Dirac neutrino
has at the two media interface and which are discussed above. So that, in
particular, there is no Majorana neutrino trapping and reflection by matter. It
should be noted that the equation (\ref{Dirac_Majorana}) and the Majorana
neutrino spectrum in matter were discussed previously also in
\cite{PanPLB91-PRD92WeiKiePRD97,
BerVysBerSmiPLB87_89GiuKimLeeLamPRD92GiuKimLeeLamPRD92BerRosPLB94}.

\subsection{Flavour neutrino energy difference in matter}
Although the neutrino energy spectra corespondent to the modified
Dirac-Pauli and Dirac equations are not the same, an equal result
given by (\ref{delta_Energy}) for the energy difference $\Delta
E=E(s=-1)-E(s=+1)$ of the two neutrino helicity states can be
obtained from both of the spectra in the low matter density limit
$\alpha\frac {pm}{E_{0}^{2}} \ll 1$.

It should be also noted that for the relativistic neutrinos the
energy spectrum for the neutrino helicity states of
Eq.(\ref{Energy}) in the low density limit \footnote{This limit is
set by a huge value that is far above any realistic densities of
astrophysical media} reproduces the correct energy values for the
neutrino left-handed and right-handed chiral states:
\begin{equation}\label{E_L}
  E_{\nu_L} \approx E(s=-1)\approx E_0 +{{\tilde {G}}_F \over \sqrt{2}}n,
\end{equation}
and
\begin{equation}\label{E_R}
  E_{\nu_R} \approx E(s=-1)\approx E_0,
\end{equation}
as it should be for the active left-handed and sterile
right-handed neutrino in matter.

We should like to note, that the obtained spectra for the flavor
neutrinos of different helicities in the presence of matter
enables one to reproduce the well-known result for the energy
difference of two flavour neutrinos in matter. In order to
demonstrate this we expand the expressions for the relativistic
electron and muon neutrino energies, which are giving by
(\ref{Energy}) for the Dirac case or by (\ref{Energy_Majorana})
for the Majorana case, over $m/p\ll 1$ and get
\begin{equation}E_{\nu_e, \nu_{\mu}}^{s=-1}\approx E_0
+ 2\alpha_{\nu_e, \nu_{\mu}} m.
\end{equation}
Then the energy difference for the two active flavour neutrinos will be
\begin{equation}\Delta E =
E_{\nu_e}^{s=-1}-E_{\nu_{\mu}}^{s=-1} = \sqrt2 G_F n_e.
\end{equation}
Analogously, considering the spin-flavour oscillations $\nu_{e_L}\rightleftarrows \nu_{\mu_R}$, for
the corresponding energy difference we find:
\begin{equation}\Delta E =
E_{\nu_e}^{s=-1}-E_{\nu_{\mu}}^{s=+1} = \sqrt2 G_F \big(n_e-{1\over 2}n_n\big).
\end{equation}
These equations enable one to get the expressions for the neutrino
flavour and spin-flavour oscillation probabilities with resonance
dependences on the matter density in the complete agreement with
the results of \cite{WolPRD78MikSmiYF85,Akh88LimMar88}.

\section{Neutrino spin light in matter}

In this section we illustrate how the method based on the use of
the exact solutions of the modified Dirac equation for the
neutrino wave function can be used in the study of different
phenomena which may exist when a neutrino moves in matter. We
consider the spin light of neutrino ($SL\nu$), a new type of
electromagnetic radiation that can be produced by the Dirac
neutrino with nonzero magnetic moment while moving in the
background matter. This phenomena was first predicted and studied
within the quasi-classical theory in
\cite{LobStuPLB03_04DvoGriStuIJMPD05}. The $SL\nu$ is a quantum
phenomenon by its nature, that is why it was important to
elaborate the quantum treatment of this process
\cite{StuTerPLB05,GriStuTerPLB05,StuJPA06}.

The $SL\nu$ in matter originates from the quantum electromagnetic
transition between the different helicity states due to the
neutrino magnetic moment interactions with photons. In this
section we give the quantum theory of the effect, that is based on
the approach similar to the Furry representation in the quantum
electrodynamics which has been discussed above in Introduction.

The corresponding Feynman diagram of the process under
consideration is shown in Fig.2. The neutrino initial $\psi_{i}$
and final $\psi_{f}$ states are described by ``broad lines" that
account for the neutrino interaction with matter.
\begin{figure}[h]\label{diagram}
\begin{center}
{  \includegraphics[scale=.7]{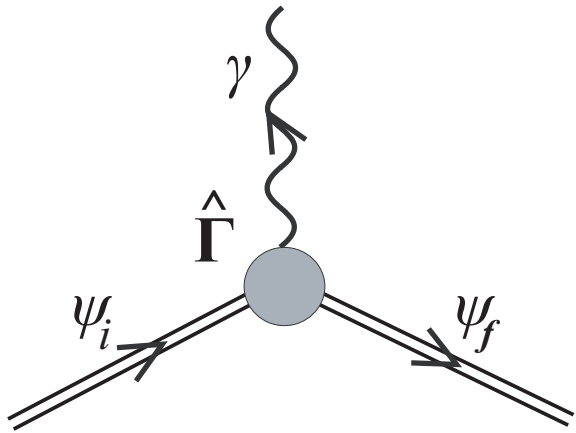}}
    \caption{
    The $SL\nu$ radiation diagram. }
  \end{center}
\end{figure}
The corresponding amplitude is given by
\begin{equation}\label{amplitude}
\begin{array}{c} \displaystyle
  S_{f i}=-\mu \sqrt{4\pi}\int d^{4} x {\bar \psi}_{f}(x)
  ({\hat {\bf \Gamma}}{\bf e}^{*})\frac{e^{ikx}}{\sqrt{2\omega L^{3}}}
   \psi_{i}(x),
   \\
   \\
   \hat {\bf \Gamma}=i\omega\big\{\big[{\bf \Sigma} \times
  {\bm \varkappa}\big]+i\gamma^{5}{\bf \Sigma}\big\},
\end{array}
\end{equation}
where $\mu$ is the neutrino magnetic moment, $k^{\mu}=(\omega,{\bf k})$ and ${\bf e}^{*}$ are the
photon momentum and polarization vectors, ${\bm \varkappa}={\bf k}/{\omega}$ is the unit vector
pointing in the direction of the emitted photon propagation. Here again we consider the case of the
electron neutrino moving in unpolarized matter composed of electrons. Then the integration over the
time and spatial coordinates in (\ref{amplitude}) gives
\begin{equation}\label{amplitude_evaluated}
   S_{f i}=
  -\mu {\sqrt {\frac {2\pi}{\omega L^{3}}}}
  ~2\pi\delta(E^{\prime}-E+\omega)\delta^3({\bf p^{\prime}}-{\bf p}+{\bf k})
  {\bar u}_{f}(E^{\prime}, {\bf p^{\prime}})({\hat {\bf \Gamma}}{\bf e}^{*})
  u_{i}(E, {\bf p}).
\end{equation}
The unprimed and primed symbols refer to initial and final neutrino states, respectively. From the
energy-momentum conservation law
\begin{equation}\label{e_m_con}
    E=E^{\prime}+\omega, \ \ \
    {\bf p}={\bf p}^{\prime}+{\bm k},
\end{equation}
it follows that photon is radiated only when the initial and final neutrino states are
characterized by $s_{i}=-1$ and $s_{f}=+1$, respectively. For the emitted photon energy we then
obtain:
\begin{equation}\label{omega1}
\omega =\frac{2\alpha mp\left[ (E-\alpha m)-\left( p+\alpha m\right) \cos \theta \right] }{\left(
E-\alpha m-p\cos \theta \right) ^{2}-\left( \alpha m\right) ^{2}},
\end{equation}
where the angle $\theta$ gives the direction of the radiation in
respect to the initial neutrino momentum $\bf p$. For the $SL\nu$
radiation rate and total power we get, respectively,
\begin{equation}
\label{Gamma}
 \Gamma =\mu ^2
 \int_{0}^{\pi }\frac{\omega ^{3}}{1+\tilde\beta ^{\prime
}y}S\sin \theta d\theta,   \ \ \
I=\mu ^2\int_{0}^{\pi }\frac{\omega ^{4}}{1+\tilde\beta
^{\prime}y}S\sin \theta d\theta,
\end{equation}
where
\begin{equation}\label{S}
S=(\tilde\beta \tilde\beta ^{\prime }+1)(1-y\cos \theta
)-(\tilde\beta +\tilde\beta ^{\prime }) (\cos \theta -y),
\end{equation}
\begin{equation}\label{beta}
\tilde \beta =\frac{p+\alpha m}{E-\alpha m}, \ \ \tilde \beta
^{\prime }=\frac{p^{\prime }-\alpha m}{E^{\prime }-\alpha m}, \ \
\ E^{\prime }=E-\omega , \ \ \ p^{\prime }=K\omega -p,
\end{equation}
\begin{equation}
y=\frac{\omega -p\cos \theta }{p^{\prime }}, \ \ K=\frac{E-\alpha
m-p\cos \theta }{\alpha m}.
\end{equation}

In the relativistic neutrino momentum case, $p\gg m$, and for
different values of the matter density parameter  $\alpha$ from
(\ref{Gamma}) and we have the following limiting values
\cite{StuTerPLB05,GriStuTerPLB05} (see also \cite{StuJPA06}):
\begin{equation}\label{p_gg}
\Gamma = \left\{
  \begin{tabular}{c}
  \ $\frac{64}{3} \mu ^2 \alpha ^3 p^2 m,$ \\
  \ $4 \mu ^2 \alpha ^2 m^2 p$, \\
  \ $4 \mu ^2 \alpha ^3 m^3$,
  \end{tabular}
\right. \ \ I= \left\{
  \begin{tabular}{cc}
  \ $\frac{128}{3}\mu ^{2}\alpha ^{4}p^{4},$ &
  \ \ \ for {$\alpha \ll \frac{m}{p},$ } \\
  \ $\frac{4}{3} \mu ^2 \alpha ^2 m^2 p^2$, & \ \ \ \ \ \ \ \
  { for
  $ \frac{m}{p} \ll \alpha \ll \frac{p}{m},$} \\
\ $4 \mu ^2 \alpha ^4 m^4$, & \ { for
  $ \alpha \gg \frac{p}{m}. $}
  \end{tabular}
\right.
\end{equation}
In the opposite case of ``non-relativistic" neutrinos, $p\ll m$,
we get:
\begin{equation}\label{p_ll}
\Gamma = \left\{
  \begin{tabular}{c}
  \ $\frac{64}{3} \mu ^2 \alpha ^3 p^3,$ \\
  \ $\frac{512}{5} \mu ^2 \alpha ^6 p^3$,\\
  \ $4 \mu ^2 \alpha ^3 m^3$,
  \end{tabular}
  \ \ \
I = \left\{
  \begin{tabular}{cc}
  \ $\frac{128}{3} \mu ^2 \alpha ^4 p^4,$ &
  \ \ \ \ \ for {$\alpha \ll 1,$ } \\
  \ $\frac{1024}{3} \mu ^2 \alpha ^8 p^4$, & \ \ \ \ \ \ \ \ \ \
  {
  for
  $ 1 \ll \alpha \ll \frac{m}{p},$} \\
  \ $4 \mu ^2 \alpha ^4 m^4$, & \ \ \ \ { for
  $ \alpha \gg \frac{m}{p}. $}
  \end{tabular}
\right. \right.
\end{equation}
It can be seen that in the case of a very dense matter the values
of the rate and total power are mainly determined by the density.
Note that the obtained above results in the case of small
densities are in agreement with the studies of the neutrino spin
light performed on the basis of the quasi-classical approach
\cite{LobStuPLB03_04DvoGriStuIJMPD05}. The $SL\nu$ characteristics
in the case of matter with ``moderate" \ densities ( the second
lines of (\ref{p_gg}) ) were also obtained in \cite{LobPLB05}.

One can estimate the average emitted photon energy $\left\langle
\omega\right\rangle = {I}/{\Gamma}$ with the use of the obtained
above values of the rate and total power (\ref{p_gg}) and
(\ref{p_ll}) for different matter densities. In the two case
($p\gg m$ and $p\ll m$), we get, respectively,
\begin{equation}\label{overage_omega}
\left\langle \omega\right\rangle \simeq \left\{
  \begin{tabular}{cc}
  \ $2\alpha \frac{p^{2}}{m},$ &
  for {$\alpha \ll \frac{m}{p},$ } \\
  \ $\frac{1}{3} p $, & { for
  $ \frac{m}{p} \ll \alpha \ll \frac{p}{m},$} \\
\ $\alpha m$, & \ { for
  $ \alpha \gg \frac{p}{m} $},
  \end{tabular}
\right.
\left\langle \omega\right\rangle \simeq\left\{
  \begin{tabular}{cc}
  \ $2 \alpha p ,$ &
  for {$\alpha \ll 1,$ } \\
  \ $\frac{10}{3}\alpha^2 p$, &
{ for
  $ 1 \ll \alpha \ll \frac{m}{p},$} \\
\ $\alpha  m $, &  { for
  $ \alpha \gg \frac{m}{p}. $}
  \end{tabular}
\right.
\end{equation}

To summaries the main properties of the spin light of neutrino in
matter, we should like to point out that this phenomenon arises
due to neutrino energy dependence in matter on the neutrino
helicity state. In media characterized by the positive values of
the parameter $\alpha$, the negative-helicity neutrinos (the
left-chiral relativistic neutrinos) are converted into the
positive-helicity neutrinos (the right-chiral relativistic
neutrinos) in the process under consideration. Thus, the neutrino
self-polarization effect can appear (see also
\cite{LobStuPLB03_04DvoGriStuIJMPD05}). From the above estimations
for the emitted photon energies it follows that for the
ultra-relativistic neutrinos moving in dense matter the $SL\nu$
can be regarded as an effective mechanism for production of the
gamma-rays.

Note that an adequate description of the $SL\nu$ in the low matter density
limit $\alpha\frac {pm}{E_{0}^{2}} \ll 1$ can be obtained on the basis of the
Dirac-Pauli equation for the neutrino wave function \cite{GriStuTer_11LomCon}.
More over, the use of the Dirac-Pauli equation enables us also to consider the
$SL\nu$ in the case of totally polarized (due to the presence of strong
magnetic field with the strength given by (\ref{str_B}) ) electron gas. For
this particular situation, the emitted $SL\nu$ photon energy, as it follows
from (\ref{redefin}), is given by
\begin{equation}\label{}
    \omega=
    \frac {\beta}{1-\beta_e \cos
    \theta}\omega_0,
\end{equation}
where
\begin{equation}\label{omega_0_1}
\omega_0= \frac {{ G}_{F}} {\sqrt{2}}n \left[ \left(1+\sin^{2}
  4\theta_{W}\right)-{sign
\left(\frac{ B_{\parallel}}{B}\right)}\right]+
    2\frac {\mu B_{\parallel}}{\gamma}.
\end{equation}

A remark on the possibility for Majorana neutrino to emit the spin
light in matter should be made. Obviously, due to the absence of
the magnetic moment, such radiation is not expected in this case.
However, having two neutrinos of different flavour, it is possible
to produce an analogous effect via the transition magnetic moment,
which Majorana neutrinos can possess.

\section{Modified Dirac equation for electron in matter}
In \cite{StuJPA06,GriShiStuTerTro_12LomCon}, it has been shown
how the approach, developed at first for description of a neutrino
motion in the background matter, can be spread for the case of an
electron propagating in matter. The modified Dirac equation for an
electron in matter has been derived \cite{StuJPA06} and on this
basis we have considered the electromagnetic radiation that can be
emitted by the electron (due to its electric charge) in the
background matter. We have termed this radiation as the \ ``spin
light of electron" in matter. It should be noted here that the
term \ ``spin light" was introduced in \cite{ITernSPU95} for
designation of the particular spin-dependent contribution to the
electron synchrotron radiation power.

Let us consider an electron having the standard model interactions with
particles of electrically neutral matter composed of neutrons, electrons and
protons. This can be used for modelling a real situation existed, for instance,
when electrons move in different astrophysical environments. We suppose that
there is a macroscopic amount of the background particles in the scale of an
electron de Broglie wave length. In fact, we account below only the neutron
component of matter. Then the addition to the electron effective interaction
Lagrangian is
\begin{equation}\label{Lag_f_e}
\Delta L^{(e)}_{eff}=-{f}^\mu \Big(\bar e \gamma_\mu
{1-4\sin^{2}\theta_{W}+\gamma^5 \over 2} e \Big),
\end{equation}
where the explicit form of $f^{\mu}$ depends on the background particles
density, speed and polarization and is determined by (\ref{Lag_f}) and
(\ref{q_f}). The modified Dirac equation for the electron wave function in
matter is \cite{StuJPA06}
\begin{equation}\label{new_e}
\Big\{ i\gamma_{\mu}\partial^{\mu}-\frac{1}{2}
\gamma_{\mu}(1-4\sin^{2}\theta_{W}+\gamma_{5}){\widetilde
f}^{\mu}-m_e \Big\}\Psi_{e}(x)=0,
\end{equation}
where for the case of electron moving in the background of
neutrons (that can be used as an abrupt model of a nuclear matter
of a neutron star, see, for instance, \cite{KusPosPLB02})
\begin{equation}
{\tilde f}^{\mu}=-f^\mu=\frac{G_F}{\sqrt
2}(j^{\mu}_n-\lambda^{\mu}_n).
\end{equation}
We consider below unpolarized neutrons so that
\begin{equation}\label{f2_e}
\widetilde{f}^{\mu}=\frac{G_{F}}{\sqrt{2}}(n_n,n_n{\bf v}),
\end{equation}
here $n_n$ is the neutrons number density and $\mathbf v$ is the
speed of the reference frame in which the mean momentum of the
neutrons is zero.

The corresponding electron energy spectrum in the case of
unpolarized matter at rest is given by
\begin{equation}\label{Energy_e}
  E_{\varepsilon}^{(e)}=
  \varepsilon \sqrt{{{\bf p}_e}^{2}\Big(1-s_e\alpha_n
  \frac{m_e}{p_e}\Big)^{2}
  +{m_e}^2} +c{\alpha}_n m_e, \ \ \alpha_n=\frac{1}{2\sqrt{2}}
  {G_F}\frac{n_n}{m_e},
\end{equation}
where $c=1-4\sin^2 \theta_W$ and  the notations for the electron
mass, momentum, helicity and sign of energy are similar to those
used in Section 2 for the case of neutrino. For the wave function
of the electron moving in nuclear matter we get \cite{
GriShiStuTerTro_12LomCon}
\begin{equation}
\Psi_{\varepsilon, {\bf p},s}({\bf r},t)=\frac{e^{-i(
E^{(e)}_{\varepsilon}t-{\bf p}{\bf r})}}{2L^{\frac{3}{2}}}
\left(%
\begin{array}{c}{\sqrt{1+ \frac{m_e}{ E^{(e)}_{\varepsilon}-c\alpha_n m_e}}}
\ \sqrt{1+s\frac{p_{3}}{p}}
\\
{s \sqrt{1+ \frac{m_e}{ E^{(e)}_{\varepsilon}-c\alpha_n m_e}}} \
\sqrt{1-s\frac{p_{3}}{p}}\ \ e^{i\delta}
\\
{  s\varepsilon\sqrt{1- \frac{m_e}{
E^{(e)}_{\varepsilon}-c\alpha_n m_e}}} \ \sqrt{1+s\frac{p_{3}}{p}}
\\
{\varepsilon\sqrt{1- \frac{m_e}{ E^{(e)}_{\varepsilon}-c\alpha_n
m_e}}} \ \ \sqrt{1-s\frac{p_{3}}{p}}\ e^{i\delta}
\end{array}%
\right)
\end{equation}
The exact solutions of this equation open a new method for investigation of
different quantum processes which can appear when electrons propagate in
matter. On this basis, we predict and investigate the $SLe$
\cite{StuJPA06,GriShiStuTerTro_12LomCon}.
 The $SLe$ photon energy,
obtained from the energy conservation law, is given by
\begin{equation}\label{omega1_e}
\omega_{SLe} =\frac{2\alpha_n m_ep_e\left[ (E^{(e)}-c\alpha_n
m_e)-\left( p_e+\alpha_n m_e\right) \cos \theta_{SLe} \right]
}{\left( E^{(e)}-c\alpha_n m_e-p_e\cos \theta_{SLe} \right)
^{2}-\left( \alpha_n m_e\right) ^{2}},
\end{equation}
where the angle $\theta_{SLe}$ gives the direction of the
radiation in respect to the initial electron momentum ${\bf p}_e$.
In the case of relativistic electrons and small values of the
matter density parameter $\alpha_n$ the photon energy is
\begin{equation}\label{omega_2}
    \omega_{SLe}=
    \frac {\beta_e}{1-\beta_e \cos
    \theta_{SLe}}\omega_0,\ \ \
\omega_0= \frac {G_{F}} {\sqrt{2}}n_n,
\end{equation}
here $\beta_e$ is the electron speed in vacuum. From this
expressions we conclude that for the relativistic electrons the
energy range of the $SLe$ may even extend up to energies peculiar
to the spectrum of gamma-rays. We also predict the existence of
the electron-spin polarization effect in this process. Finally,
from the order-of-magnitude estimation, we expect that the ratio
of rates of the $SLe$ and the $SL\nu$ in matter is
\begin{equation}
R=\frac {\Gamma_{SLe}}{\Gamma_{SL\nu}}\sim \frac {e^2}{\omega^2
\mu^2},
\end{equation}
that gives $R \sim 10^{18}$ for the radiation in the range of gamma-rays,
$\omega \sim 5 \ MeV$, and for the neutrino magnetic moment $\mu \sim
10^{-10}\mu_0$. Thus, we expect that in certain cases the $SLe$ in matter would
be more effective than the $SL\nu$.

\section{Conclusion}

In this paper, we have developed a framework for treating
different interactions of neutrinos and electrons in the presence
of matter. This method is based on the use of modified Dirac
equations for particles wave functions, in which the correspondent
effective potentials that account for the interaction with matter
are included.

For a neutrino moving in matter, we have considered the modified
Dirac-Pauli and Dirac equations and evaluated the correspondent
wave functions and energy spectra in the presence of matter. Both
cases of Dirac and Majorana neutrinos have been discussed.

Within the framework developed, we have also derived the modified
Dirac equation for the electron moving in matter and evaluated the
exact wave functions and energy spectra.

The approach developed is similar to the Furry representation
which is used in quantum electrodynamics in investigations of
particles interactions in the presence of external electromagnetic
fields. Note that our focus has been on the standard model
interactions of neutrinos and electrons with the background
matter. The same approach, which implies the use of the exact
solutions of the correspondent modified Dirac equations, can be
used in the case when neutrinos and electrons interact with
different external fields predicted within extensions of the
standard model (see, for instance, \cite{CollKosPRD98,
ZhuLobMurPRD06}).

In conclusion, we should like to note that the approach developed
is valid in the case when the interaction of neutrinos and
electrons with particle of the background is coherent. This
condition is satisfied when a macroscopic amount of the background
particles are confined within the scale of a neutrino or electron
de Broglie wave length. So that for the relativistic neutrinos and
electrons $(l=\nu$ or $e )$ the following condition must be
satisfied
\begin{equation}\label{d_Br}
\frac {n}{\gamma_l m^3_l}\gg 1,
\end{equation}
where $n$ is the number density of matter and
$\gamma_l=\frac{E_l}{m_l}$. For instance, let us consider the case
of neutrino. If we express $n$ by the dimensional number $N$
following to $n=N \ cm^{-3}=N\times 2^3\times 10^{-15}eV ^3$ and
for the neutrino mass use the value of $m_\nu \sim 1\ eV$, then
from (\ref{d_Br}) we have
\begin{equation}\label{d_Br_1}
N\gg  10^{14}\times \Big( \frac{E_\nu}{1\ eV} \Big).
\end{equation}
It follows that even for not extremely dense astrophysical matter
with $N\sim 10^{33}$ (this value is about five orders of magnitude
lower then one peculiar to densities of neutron stars) the
approach developed is valid for the neutrino ultra-high energy
band.

\end{document}